\documentstyle[twocolumn,short,epsf]{jpsj}
\newcommand{\simj}{\stackrel{>}{_\sim}}
\newcommand{\simk}{\stackrel{<}{_\sim}}
\def\runauthor{K. {\sc Sano} and K. {\sc Takano}}
\title{Spin Gap of $S=1/2$ Heisenberg Model \\
on Distorted Diamond Chain
}
\author
{ 
Kazuhiro {\sc Sano}
and Ken'ichi {\sc Takano}$^{1}$
}
\inst
{
Department of Physics Engineering, \\
Mie University, Tsu, Mie 514-8507 \\
$^1$Laboratory of Theoretical Condensed Matter Physics and \\ 
Research Center for Advanced Photon Technology, \\
Toyota Technological Institute, Nagoya 468-8511 \\
}
\recdate{\hskip 5cm}
\kword
{
spin gap, one-dimensional Heisenberg model, 
distorted diamond chain, frustration, 
numerical diagonalization, 
$\rm Cu_3 Cl_6 (H_2 O)_2 \cdot 2H_8 C_4 SO_2$ 
}
\begin{document}
\sloppy
\maketitle
    Recently, Ishii et al.\cite{Ishii} measured 
the magnetic susceptibility $\chi$ for 
$\rm Cu_3 Cl_6 (H_2 O)_2 \cdot 2H_8 C_4 SO_2$, which is 
considered to be a quasi-one-dimensional material consisting 
of $S=\frac{1}{2}$ trimer spin chains. 
    The result indicates that $\chi$ vanishes in the low 
temperature limit. 
    They also measured the magnetization process for this 
material, and showed that there is a plateau of zero 
magnetization below the critical field $H_c \simeq 3.9$ T. 
    From these experimental results, they concluded that the 
ground state is a singlet state with spin gap. 
    The spin gap $\Delta$ is estimated as 
$\Delta/k_{\rm B} \simeq 5.2$ K from the value of $H_c$. 

\begin{figure}[btp]
\vspace{0.4cm}
\begin{center}\leavevmode
\epsfxsize=70mm
\epsfbox{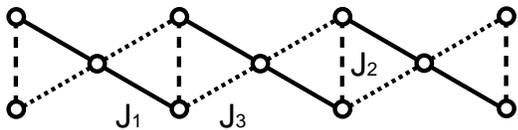}
\end{center}
\vspace{-0.6cm}
\caption{
    The lattice structure of the distorted diamond chain 
representing $\rm Cu_3 Cl_6 (H_2 O)_2 \cdot 2H_8 C_4 SO_2$. 
    Three kinds of exchange constants are shown. 
}
\label{fig1}
\end{figure}

    The proposed Hamiltonian\cite{Ishii} representing a spin chain 
in this material is given by 
\begin{eqnarray}
   H&=&J_1 \sum_j \left( {\mib S}_{3j-1} \cdot {\mib S}_{3j} 
     + {\mib S}_{3j} \cdot {\mib S}_{3j+1} \right) \nonumber \\
    &+&J_2 \sum_j {\mib S}_{3j+1} \cdot {\mib S}_{3j+2} \nonumber \\
    &+&J_3 \sum_j \left( {\mib S}_{3j-2} \cdot {\mib S}_{3j}
          + {\mib S}_{3j} \cdot {\mib S}_{3j+2} \right), 
\label{eq1}
\end{eqnarray}
where ${\mib S}_{j}$ is the $S=\frac{1}{2}$ spin on site $j$. 
    Three spins ${\mib S}_{3j-1}$, ${\mib S}_{3j}$ and 
${\mib S}_{3j+1}$ form a trimer. 
    The lattice structure is shown in Fig.~\ref{fig1}. 
    Three kinds of exchange constants $J_1$, $J_2$ and $J_3$ are 
inferred to be positive and to satisfy the relation 
$J_1 > J_2, J_3$ from the lattice parameters of the material 
\cite{Swank}. 
    Hereafter, we use the unit of $J_1 = 1$. 

    The symmetric case of $J_3 = J_1 (=1)$ has been studied by 
Takano et al.\cite{Takano} and the system has been called the 
{\it diamond chain}. 
    They almost exactly showed that there exist three phases in 
the parameter space; the ferrimagnetic phase for $J_2 < 0.909$, 
the tetramer-dimer (TD) singlet phase for $0.909 < J_2 < 2$ 
and the dimer-monomer (DM) singlet phase for $ J_2 > 2$. 
    The TD phase is a disordered phase with spin gap which 
originates from frustration among exchange interactions, 
while the DM phase is a spin fluid phase without spin gap. 
    Okamoto et al.\cite{Okamoto} studied the general case of 
$J_3 \neq 1$; i.~e. the {\it distorted diamond chain}. 
    The three phases develop in the $J_2$-$J_3$ plane. 
    They numerically determined the phase boundaries. 
    Also Tonegawa et al.\cite{Tonegawa} numerically studied the 
magnetization process and showed plateaux for $\frac{1}{3}$ and 
$\frac{2}{3}$ of the saturation field. 

    In this article, we estimate the values of the spin gap 
by the numerical diagonalization. 
    Then we produce a contour map in the $J_2$-$J_3$ parameter 
space. 
    The contour map represents an overall feature of the gapped 
phase of the $S=\frac{1}{2}$ Heisenberg model on the distorted 
diamond chain. 
    When further experimental information on 
$\rm Cu_3 Cl_6 (H_2 O)_2 \cdot 2H_8 C_4 SO_2$ is given, 
the contour map will be useful to determine the values of the 
exchange constants for the real material. 

\begin{figure}[btp]
\begin{center}\leavevmode
\epsfxsize=75mm
\epsfbox{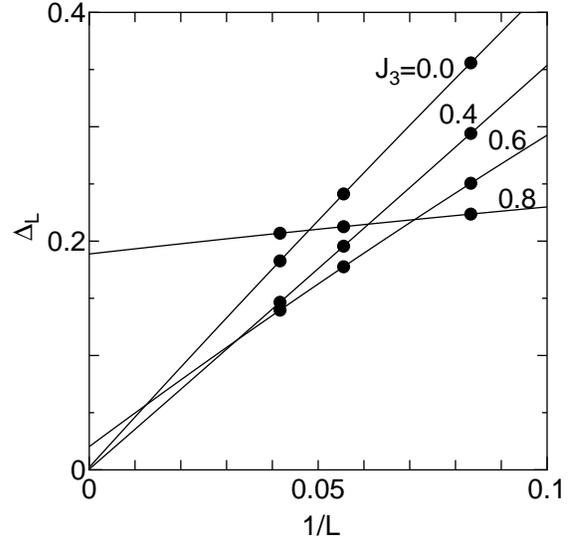}
\end{center}
\caption{
Spin gaps of finite systems with $L$ = 12, 18 and 24 (solid circles) 
for several values of $J_3$ at $J_2=1$. 
Each solid line represents eq.~(\ref{eq2}) with fitting parameters 
$c_1$, $c_2$ and $\Delta_{\infty}$ for a value of $J_3$. 
}
\label{fig2}
\end{figure}
    We first calculate the spin gap $\Delta_L$ for finite chains 
with system size $L$. 
    The spin gap $\Delta_{\infty}$ in the thermodynamic limit 
is evaluated by extrapolation.
    We assume the size dependence of $\Delta_L$ as 
\begin{equation}
    \Delta_L= \Delta_{\infty}+ \frac{c_1}{L} + \frac{c_2}{L^2} 
\label{eq2}
\end{equation}
with constants $c_1$ and $c_2$. 
    The numerical diagonalization has been done for $L=$12, 18 
and 24 under the periodic boundary condition. 
    We determine $c_1$, $c_2$ and $\Delta_{\infty}$ by fitting. 
    In Fig.~\ref{fig2}, we show $\Delta_L$ as a function of $L$ for 
several values of $J_3$ at $J_2=1$. 
    For $J_3=0$, the estimated value of $\Delta_{\infty}$ 
is about 0.002 and is close to zero; 
    the nonzero value is interpreted as an extrapolation error 
\cite{exact}. 
    For $0 < J_3 \simk 0.4$, the true value of the spin gap is 
very small or may be regarded as zero, since the estimated values 
are less than 0.002 and are within the extrapolation error. 
    For $J_3 \simj 0.6$, the figure shows that the system has 
a finite spin gap. 
    For $J_3=0.5$, $\Delta_{\infty}$ is 0.0034, 
which is small but seems to be finite. 
    This is consistent with the result of Okamoto et al. 
that the spin gap opens at the critical value 
$J_3^c \simeq 0.35$ for $J_2=1$ \cite{Okamoto}. 

\begin{figure}[btp]
\begin{center}\leavevmode
\epsfxsize=75mm
\epsfbox{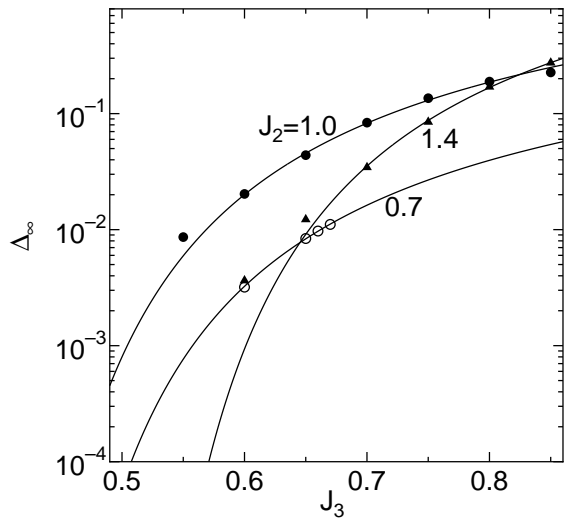}
\end{center}
\caption{
The spin gap $\Delta_{\infty}$ in the thermodynamic limit 
(symbols) as a function of $J_3$ for $J_2$ = 0.7, 1.0 and 1.4. 
Each solid line represents $D(J_3)$ (eq.~(\ref{eq3})) 
with fitting parameters 
$a_1$ and $a_2$ for a value of $J_2$. 
}
\label{fig3}
\end{figure}

    In general, the spin gap in a dimer phase is exponentially 
small near the phase boundary to a spin fluid phase. 
    Hence it is difficult to estimate $\Delta_{\infty}$  
near the boundary in the present case. 
    To overcome this difficulty, we assume that $J_3$ dependence
 of the spin gap is given by 
\begin{equation}
    D(J_3) = a_1 \sqrt{J_3-J_3^c} 
\exp\left( -\frac{a_2}{J_3-J_3^c} \right) 
\label{eq3}
\end{equation}
for $J_3 \sim J_3^c$\cite{Haldane}, where $a_1$ and $a_2$ are 
constants. 
    We have the values of $J_3^c$ 
by inspecting the phase diagram of Okamoto et al. \cite{Okamoto}; 
e.~g. $J_3^c$ = 0.374, 0.354 and 0.460 
for $J_2$ = 0.7, 1.0 and 1.4, respectively. 
    We carry out the fitting of the extrapolation data 
$\Delta_{\infty}$ to eq.~(\ref{eq3}) and determine $a_1$ and $a_2$. 
    Figure \ref{fig3} represents the fitting function 
$D(J_3)$ and the extrapolation data. 
    We find that the extrapolation data are well reproduced by 
eq.~(\ref{eq3}) for $\Delta_{\infty} \simj 0.02$. 
    Hence the function form in eq.~(\ref{eq3}) is reliable. 
    We use eq.~(\ref{eq3}) to estimate the spin gap for 
$\Delta_{\infty} \simk 0.02$ near the critical value $J_3^c$. 
    For example, the spin gap is estimated as $1.0 \times 10^{-2}$ 
at $J_3 \simeq 0.57$, $1.0 \times 10^{-3}$ at $J_3 \simeq 0.50$, 
$1.0 \times 10^{-4}$ at $J_3 \simeq 0.47$ and $1.0 \times 10^{-5}$ 
at $J_3 \simeq 0.45$ for $J_2=1.0$. 

\begin{figure}[btp]
\begin{center}\leavevmode
\epsfxsize=75mm
\epsfbox{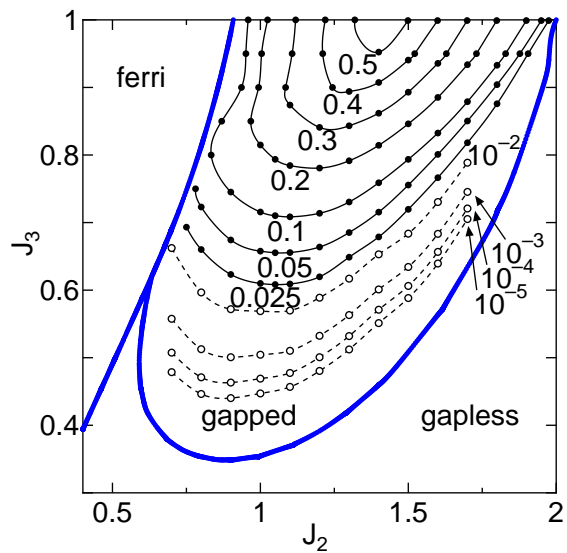}
\end{center}
\caption{
    The contour map for the spin gap in the gapped phase of 
the $J_2$-$J_3$ plane. 
    Bold solid lines are the phase boundaries that Okamoto et al. 
have determined \cite{Okamoto}. 
}
\label{fig4}
\end{figure}

    Using these results, we draw contour lines of the spin gap 
in the $J_2$-$J_3$ plain. 
    The resultant contour map is shown in Fig.~\ref{fig4}. 
    We have calculated $\Delta_{\infty}$ at the discrete positions 
($J_2$, $J_3$) with $J_2$ = 0.7, 0.8, ..., 2.0 and 
$J_3$ = 0.5, 0.55, ..., 1.0. 
    For $\Delta_{\infty} > 0.02$, the positions 
of solid circles are determined by the linear interpolation 
among the spin gaps $\Delta_{\infty}$ at the discrete positions. 
    For $\Delta_{\infty} < 0.02$, the positions of
open circles are determined by using $D(J_3)$ (eq.~(\ref{eq3})) 
instead of $\Delta_{\infty}$. 

    The temperature dependence of the experimental magnetic 
susceptibility has a broad peak at $\sim$70 K. 
    It suggests that the  energy scale of the characteristic 
exchange constant $J_1$ is larger than 70 K.
    Here  we consider a case of $J_1$ being 100 K as an 
example.\cite{J1} 
    In this case, we have $\Delta_{\infty} \sim 0.05$ 
according to the observed spin gap $\sim$5 K.
    Then $J_2$ and $J_3$ are limited 
to values close to the contour line of $\Delta_{\infty}=0.05$ 
and of $J_2 < 1$. 

    One of the authors (K. T.) would like to thank 
H. Tanaka and M. Ishii for explaining their experimental results, 
and K. Okamoto for discussion. 
    This work is partially supported by the Grant-in-Aid for 
Scientific Research from the Ministry of Education, Science, 
Sports and Culture, Japan.



\begin{thebibliography}{99}
%
\bibitem{Ishii}
M.  Ishii, H. Tanaka, M. Mori, H. Uekusa, Y. Ohashi, K.  Tatani, 
Y.  Narumi and K. Kindo: 
J. Phys. Soc. Jpn. {\bf 69} (1999) 340. 
\bibitem{Swank}
 D. D. Swank and R. D.  Willett: 
Inorg. Chimica Acta {\bf 8}  (1974) 143. 
%
\bibitem{Takano}
K.  Takano, K. Kubo  and H. Sakamoto: 
J. Phys. :Condens. Matter   {\bf 8} (1996) 6405. 
%
\bibitem{Okamoto}
K. Okamoto, T. Tonegawa, Y. Takahashi and M. Kaburagi: 
J. Phys.: Condens. Matter {\bf 11} (1999) 10485. 
%
\bibitem{Tonegawa}
T. Tonegawa, K. Okamoto, T. Hikihara, Y. Takahashi and M. Kaburagi: 
 cond-mat/9912482 (1999). 
%
\bibitem{exact}
In this case, the Bethe Ansatz solution exactly shows that the 
system has a gapless excitation. 
%
\bibitem{Haldane}
F. D. M. Haldane: Phys. Rev. {\bf B25} (1982) 4925. 
%
\bibitem{J1} 
For example, 
the peak of the magnetic susceptibility is located at $\sim$0.6$J$ 
for the spin chain with uniform nearest-neighbor interactions 
(J. C. Bonner and M. E. Fisher: Phys. Rev. {\bf 135} (1964)  A640), 
and at $\sim$0.7$J$ for the two-leg spin ladder 
and the spin chain with bond alternation. 
(T. Barnes and J. Riera: Phys. Rev. {\bf B50} (1994) 6816). 
%
\end{thebibliography}
\end{document}